# Fatigue crack growth in bearing steel under cyclic mode II + static biaxial compression

Mael Zaid[1,2,3], Vincent Bonnand[1], Véronique Doquet[3], Vincent Chiaruttini[1], Didier Pacou[1], Pierre Depouhon[2]

[1]*Université Paris-Saclay, ONERA, Matériaux et Structures, 92322, Châtillon, France*

[2]*Airbus Helicopters, Aéroport International Marseille Provence, 13700 Marignane*

[3]*Laboratoire de Mécanique des Solides, CNRS, UMR 7649, Ecole Polytechnique, Institut Polytechnique de Paris, France*

Corresponding author: vincent.bonnand@onera.fr

**Abstract**: Mode II fatigue crack growth under reversed shear and static biaxial compression was investigated in two bearing steels. Many aborted branches, quasi-orthogonal to the main crack, were observed along the crack face. The compressive stress parallel to the main crack hindered the growth of these branches and favored coplanar mode II crack growth. The crack face sliding displacement profiles measured by DIC were used to derive $\Delta K_{II,eff,}$ at the main crack tip, using elastic-plastic FE simulations with crack face friction, by an inverse method. Friction-corrected crack growth kinetics were obtained for mode II crack growth in both steels.

# I. Introduction

Following the first works of Way [1], rolling contact fatigue (RCF) has been the subject of numerous studies [2–5]. However, the damage process, and especially crack growth are not fully understood. RCF is characterized by its complex stress components (hertzian stress), a negative stress triaxiality, high gradients, and non-proportional mixed-mode loading that it induces[3,6–10]. Classical bifurcation criteria do not allow a reliable prediction of the crack path for this kind of loading [11–15]. Theoretical or numerical works on the influence of biaxial compression on the bifurcation of a closed crack loaded in mode II can be found in the literature, among which those of Melin [16], Frelat and Leblond [17], Li et al. [18]. However, these works, which analyze static mode II, and not reversed cyclic shearing with static compression, do not agree regarding the impact of compression on the kink angle. They focus on crack kinking from the tip, while -as shown below- branches also develop from the crack flanks, and compete with the main crack, especially when compression is present. Besides, crack face friction in these works is taken into account using Coulomb's law, with a constant friction coefficient, which, as observed in this work and in others [19,20], does not capture the effects of normal compression on the tribological phenomena. In RCF of bearing parts, two stages can be distinguished:

  i. Crack initiation occurs at the bearing surface. The crack first propagates in mode II, because of the subsurface cyclic shearing stress, while the compressive stress field hinders mode I crack growth, and induces a high friction between crack faces. Numerous branches initiate at 90-110° along the main crack flanks. A competition between the growth of these branches, which can induce spalling if they reach the free surface, and that of the main crack is observed.
  ii. After reaching a few millimeters deep, the crack bifurcates, to grow mostly in mode I until failure.

This paper will focus on the first stage, just after crack initiation, that is: shear-driven coplanar growth in a compressive stress field.

Many papers [21–25] show crack propagation in mode II over substantial lengths in metals, while classical criteria predict immediate bifurcation [26–28]. However, it proves challenging to reproduce the growth of a crack subjected to RCF conditions in laboratory experiments [29–31]. Indeed, the most popular tests carried out in mixed-mode do not allow a coplanar propagation of the crack [14]. In order to determine mode II crack growth kinetics, Lenkovskiy [32] investigated numerous experimental methods, and none enabled coplanar propagation under mode II cyclic loading. The main issue was to avoid any tensile stress in the vicinity of the crack tip and to load it in pure shear mode. The use of compact tension-shear (CTS) specimens allowed other teams to observe coplanar crack growth in mode II over distances that increased with the loading range in various steels, and a Titanium alloy [23–25]. Otsuka et al. [21] superimposed a static compressive stress (also called non-singular "T stress") parallel to the crack submitted to cyclic shearing, and succeeded in obtaining coplanar growth in pure mode II in aluminum alloys.

In mode II, the effects of various extrinsic phenomena (crack face contact and friction, oxidation, plastic deformation of asperities, wear and debris formation) make the analysis of the crack growth kinetics and path complex [5,23]. These interactions decrease the effective range of the stress intensity factor in mode II ($\Delta K_{II,eff}$), even without any normal compression, because of asperities along the crack faces, which oppose the sliding displacements [23,33]. An experimental method has been proposed by Smith & Smith [23] to take into account these effects, and deduce $\Delta K_{II,eff}$ from the measured crack sliding displacement. This method has later been improved by Bertolino & Doquet [33] to separate, through numerical simulations, the opposite effects of crack face friction and crack tip plasticity on the measured sliding displacements, and get a more accurate estimate of $\Delta K_{II,eff}$. More recently, Bonniot et al. [34] improved the method further, to overcome the limitations associated with the assumption of stress-free crack faces in the traditional post-treatment of Digital Image Correlation (DIC) data to determine $\Delta K_{II,eff}$. A normal compressive stress reduces the effective $\Delta K_{II}$ because of an enhanced friction between the crack lips [14,23], and thus reduces the mode II crack growth rate. Similarly, for a crack subjected to cyclic mode III, Brown & al [35], as well as Tschegg and Stanzl [36] demonstrated that a static normal compressive stress increases crack face friction and induces a reduction of the crack growth rate. But until now, experimental data on the effect of a static biaxial compression, like the one encountered in RCF, on mode II crack growth are quite limited or even missing, and this case, will thus be investigated in this paper.

An additional complexity is added since many authors [25,30,31,37] have noted the presence of secondary branches that initiate quasi normal to the flanks of the mode II crack. The interactions between the coplanar crack and all the secondary branches, especially through the shielding effect, influence the crack path and growth rate. The presence of a static biaxial compression is expected to influence this phenomenon as well.

The purpose of this paper is first, to provide an experimental methodology for shear mode fatigue crack growth tests under an additional constant biaxial compression and second, to improve the understanding of mode II crack growth in conditions representative of RCF. To this end, tests on bearing steel cruciform specimens were performed at different shear loading ranges and static biaxial-compression. The near tip displacement fields were measured by DIC. A numerical model with an elastic-plastic constitutive model was developed to estimate the effective stress intensity factor allowing to establish a Paris law in pure mode II through the inverse method developed in [33,34]. This model takes into account a uniform frictional contact between the crack lips, and provides information regarding the evolution of an apparent crack face friction coefficient, discussed in relation with observations of the fracture surfaces and its varying degree of oxidation. Finally, the competitive growth of the secondary near-transverse branches and the main crack is simulated, in order to explain the continued coplanar growth in presence of biaxial compression.

## II. Experimental and numerical methods

This section describes all experimental and numerical details regarding the tests campaign reported in this paper. The tests were performed mostly on 16NCD13 bearing steel. One test was also run on 58NCD13 steel, representative of the carburized outer layer of bearing structures. The chemical compositions of the two steels are indicated in table 1. 16NCD13 steel has a bainitic/martensitic microstructure. The size of the lamellae colonies sharing the same crystallographic orientation is $65 \pm 23.3$ µm. It is not textured. 58NCD13 steel is also bainitic/martensitic, with many carbides. Elastic-plastic constitutive equations were identified, for both steels, based on data from tension-compression tests (Appendix). These equations were necessary to estimate effective stress intensity factors, as it will be discussed later.

Table 1 : nominal composition of the steels (weight %)

|         | C    | Si       | Mn      | S      | P      | Cr      | Ni      | Mo      | Fe      |
|---------|------|----------|---------|--------|--------|---------|---------|---------|---------|
| 16NCD13 | 0.16 | 0.28     | 0.48    | < 0.01 | < 0.02 | 0.96    | 3.14    | 0.23    | Balance |
| 58NCD13 | 0.58 | 0.15-0.4 | 0.3-0.6 | 0.01   | 0.015  | 0.8-1.1 | 3.0-3.5 | 0.2-0.3 | Balance |

Cruciform specimens (figure 1) were used for the assessment of the fatigue crack growth under cyclic mode II with superimposed static biaxial-compression. The specimen was designed to get a homogeneous stress distribution in the central area (radius of 20 mm and thickness of 1.5 mm), where a 3 mm-long, 200 µm-wide notch, with either rectangular or circular ends, with a radius of 100 µm was cut with an inclination of 45° relative to the tension/compression loading directions.

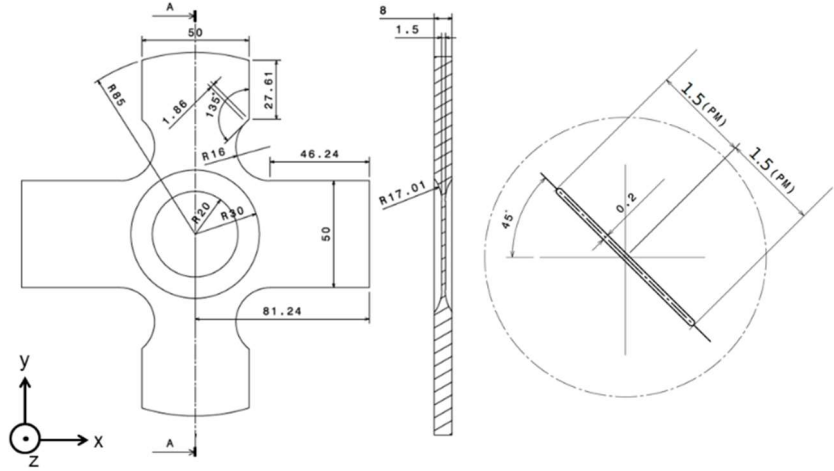

**Figure 1: Test specimen geometry**

The cruciform specimens were mounted on a biaxial servohydraulic tension-compression machine with four actuators of 250 kN capacity. No precracking was made for the tests on 16NCD13 steel, while a precracking in mode I was performed on 58NCD13 steel, to initiate approximately 1mm long cracks from both notch roots.

In order to determine the loading conditions for shear-mode loading plus biaxial compression, a preliminary FE elastic computation (figure 2) is performed with a uniform pressure applied on the end of the four "arms" of a notch and crack-free cruciform specimen. This step allows to relate the components $\sigma_{xx}$ and $\sigma_{yy}$ of the stress tensor (in the reference frame drawn on figure 2) at the center of the specimen, to the applied loads $F_x$ and $F_y$, as:

$$\sigma_{xx} = a.F_x + b.F_y, \quad (1)$$

$$\sigma_{yy} = c.F_x + d.F_y, \quad (2)$$

where a, b, c and d are geometry-dependent coefficients. Once this calibration is completed, the stress tensor in the notch coordinates system can be expressed as a function of the applied loads, and the inversion of these equations allows the computation of the time evolution of $F_x$ and $F_y$ necessary to submit a crack to a fully reversed cyclic shear stress, with a superimposed static biaxial compression, as shown in Figure 2. The shear stress $\tau$ and the normal compression $\sigma_{comp}$ can be expressed as:

$$\tau = \frac{\sigma_{yy} - \sigma_{xx}}{2} \quad (3)$$

$$\sigma_{comp} = \frac{\sigma_{xx} + \sigma_{yy}}{2} \quad (4)$$

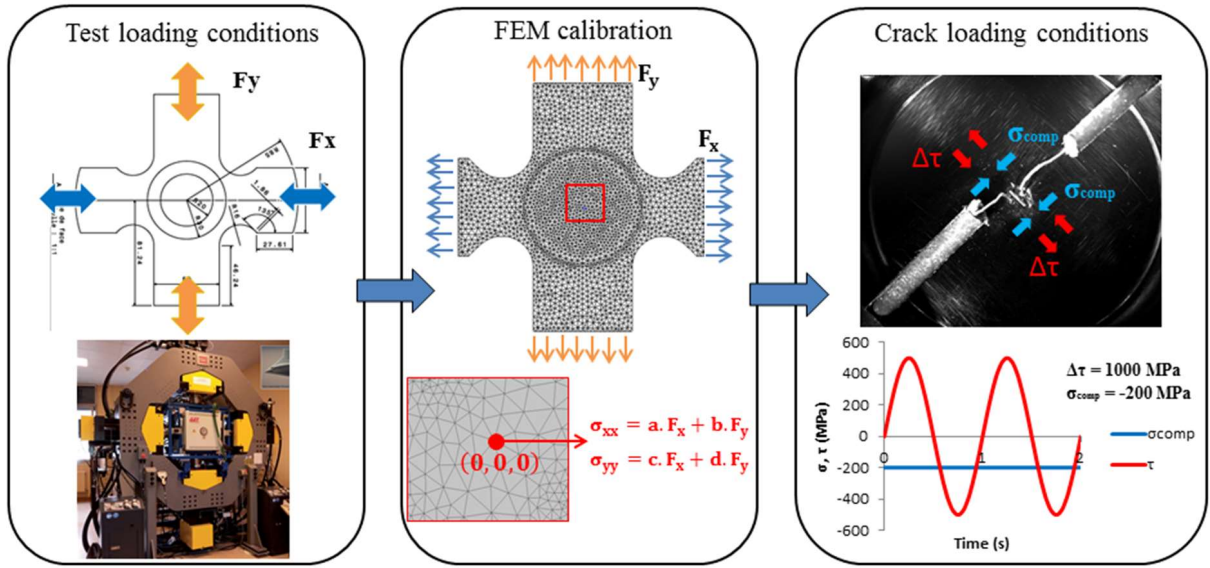

**Figure 2: a-b) Applied loading conditions in the machine coordinate axes, and c) resulting stresses in the notch plane**

Three tests were performed on 16NCD13 steel at 2 Hz, with different loading conditions, summarized in Table 2, which also reports the resulting crack paths, discussed later:

- Test #1: The purpose of this preliminary test was to find which loading function must be applied to obtain mode II crack growth, without bifurcation. A fully reversed shear loading was applied, and its amplitude was reduced stepwise, while monitoring the crack path. Then, various combinations of reversed shear and biaxial compression were applied to observe the resulting crack path.

- Test #2: One of the combinations of cyclic shear and biaxial compression, found to induce mode II crack growth during test #1, was applied during the whole test.

- Test #3: The purpose of this test was to investigate the lower part of the Paris curve in mode II by continuously decreasing the shear stress range.

- Test #4 was performed on 58NCD13 steel at 5 Hz, with loading conditions also summarized in Table 2, to grow a crack in pure II mode at relatively high propagation rate.

Table 2: Tests conditions and resulting crack paths. The coplanar crack growth was measured on the longest crack (Crack #1 for each test presented on figure 6)

| | Loading Sequence | Δτ (MPa) | Stress along x axis (MPa) | Stress along y axis (MPa) | Coplanar growth (mm) | Number of Cycles |
|---|---|---|---|---|---|---|
| Test #1 16NCD13 | S1 | 440 | 0 | 0 | 0 | 27502 |
| | S2 | 400 | 0 | 0 | 0 | 66960 |
| | S3 | 350 | 0 | 0 | 0 | 4543 |
| | S 4 | 280 | 0 | 0 | 0 | 140879 |
| | S 5 | 200 | 0 | 0 | 0 | 1772368 |
| | S 6 | 350 | -280 | -280 | 0 | 126497 |
| | S 7 | 450 | -325 | -325 | 0 | 347960 |
| | S 8 | 550 | -275 | -275 | 0 | 2569487 |
| | S 9 | 1000 | -200 | -200 | 6.1 | 8670 |
| | S 10 | 800 | -200 | -200 | 3.5 | 2090 |
| | S 11 | 600 | -200 | -200 | 0 | 57878 |
| | S12 | 800 | -200 | -200 | 5.7 | 10635 |
| Test #2 16NCD13 | | 1000 | -200 | -200 | 19.8 | 1223 |
| Test #3 16NCD13 | S 1 | 800 | -200 | -200 | 0 | 16757 |
| | S 2 | 900 | -200 | -200 | 6.8 | 5056 |
| | S 3 | 900 → 90 within 5000 cycles | -200 | -200 | 2.3 | 5020 |
| | S 4 | 900 → 540 within 5000 cycles | -200 | -200 | 12.5 | 5170 |
| Test #4 58NCD13 | S 1 | 1400 | -300 | -300 | 0.2 | 49 |
| | S 2 | 1600 | -400 | -400 | 0.19 | 27 |
| | S 3 | 0 | +450 | +450 | 0 | 316 |
| | S 4 | 1600 | -600 | -400 | 2.3 | 274 |
| | S 5 | 1600 | -400 | -600 | 0.3 | 20 |
| | S 6 | 1600 → 1130 within 650 cycles | -600 | -400 | 2.2 | 649 |
| | S 7 | 1600 → 1390 within 875 cycles | -600 | -400 | 6.6 | 875 |

To monitor crack growth, direct optical observation was made over a 40.5 x 34.2 mm-large observation field, with a spatial resolution of 16 µm/pixel. The central area of the specimen, on the side where the optical tracking was made was polished to achieve a roughness Ra = 0.05, and allow a measurement of crack path and length, based on optical contrast, enhanced by 4 low-angled LED lighting sources to produce a dark-field illumination. Digital Images Correlation (DIC) was set up to measure the displacement field over a 83.5 x 61.3 mm-large field, covering the central area of the specimen, and more specifically, to measure the crack opening and sliding displacements. A speckle pattern was first applied with sprayed black and white paints on one face of the specimen, to induce the necessary contrast for DIC. A spatial resolution of 35.5 µm/pixel was achieved. In order to determine the effective stress intensity factor $\Delta K_{II,eff}$, image pairs were captured at minimum and maximum shear stress, every other 500 cycles for test #1, every other 10 cycles for test #2 and test #4, and every other 100 cycles for test #3.

Since the contacts on the crack lips make an identification based on Williams series solution inaccurate, the crack tip position was deduced from the measured displacement field, an approach based on the evolution of the displacement gradient standard deviation along a potential crack path was used [39,40]. The map of the maximum gradient of the displacement norm, $g_m$, is chosen as an accuracy indicator of the estimated crack path:

$$g_m = \max(\|\underline{u}_{i,j+1} - \underline{u}_{i,j}\|, \|\underline{u}_{i+1,j} - \underline{u}_{i,j}\|, \|\underline{u}_{i+1,j+1} - \underline{u}_{i,j}\|, \|\underline{u}_{i-1,j+1} - \underline{u}_{i,j}\|), \quad (5)$$

where (i,j) designate respectively the line and column position of a pixel in an image. The crack indeed corresponds to a high and localized displacement gradient (figure 3(a)). A crack path tracking algorithm [39] is used, inspired by ridge tracking algorithms, to obtain the crack path by approximating the ridge line of a scalar field that presents a local extremum at the crack lips. To detect the crack tip location, the criterion proposed by Feld-Payet & al. [40] uses the local normalized standard deviation, which tends towards a threshold value near a crack tip (figure 3(b)). This threshold can be non-zero due to localized plasticity or heterogeneities of the material. The results from this approach were compared with those from the direct optical observations and a good agreement on crack tip location was observed (error of about 2% on the measured crack length).

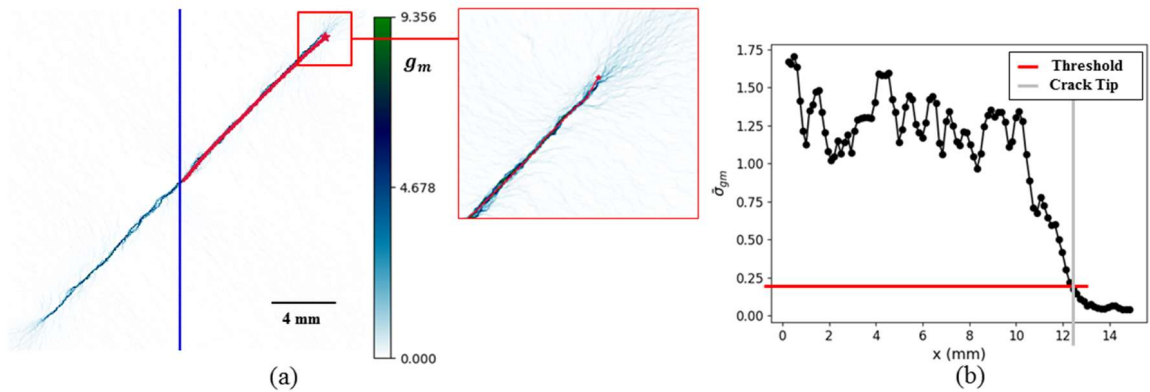

Figure 3: Crack tip detection: (a) Crack location thanks to map of $g_m$, (b) Local normalized standard deviation evolution along the crack line.

In shear mode, many interactions between crack lips (contact, friction, interlocking, asperities, wear) influence both crack growth rate and crack path, and it is thus essential to take them into account when deriving the Paris law. As mentioned by various authors [19,30,35], the nominal stress intensity factor, $\Delta K_{II}$ computed for a smooth, frictionless crack over-estimates the effective crack driving force, $\Delta K_{II,eff}$, which must be determined, based on experimental data. To that end, the classical method based on a comparison of measured displacement fields with those predicted by Williams' series expansion [41] could not be used, because the assumption of stress-free crack faces was not valid in these tests. An inverse method considering crack-tip plasticity (which tends to make crack sliding displacement larger than predicted by Linear Elastic Fracture Mechanics), as well as the contact and friction stresses along the crack faces (which, at the contrary, tends to reduce it) was thus used. First, a Crack Sliding Displacement range (ΔCSD) profile, based on the correlation of two images captured at maximum and minimum shear load, is measured, using two rows of virtual extensometers located at δ = 0.6 mm above and below the crack (figure 4).

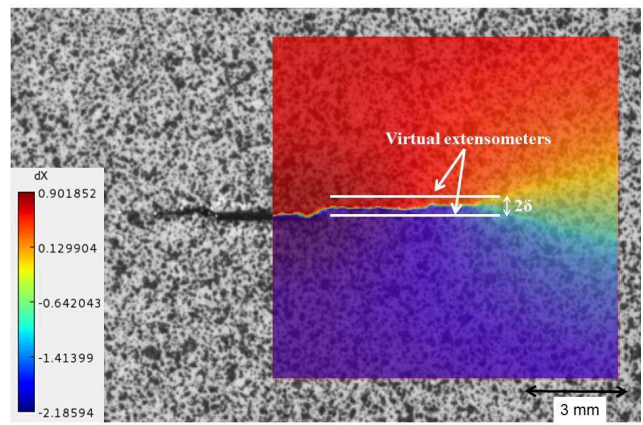

Figure 4: In-plane sliding displacement (in pixel) obtained by the DIC Software Escale [42]

Then, an elastic-plastic FE computation with frictional contact, using the constitutive equations presented in Appendix is run to capture the adverse influences of crack tip plasticity and crack face friction on the amplitude of the sliding displacement. A uniform Coulomb friction law is used, with various friction coefficients, μ. The numerical model is a 3D cruciform specimen meshed with T3 (linear triangular) elements. A refined mesh along crack faces is constructed, in order to accurately describe the contact behaviour. The size of the smallest element is 0.0375 mm at the crack tip, to get an accurate estimation of SIFs. The initial crack is inserted into the model using the adaptative meshing features developed in the *Z-Cracks* software [43]. A cyclic pressure which reproduces the experimental loading is imposed at the end of the four "arms". Finally, the ΔCSD profile computed as the CSD jump between two rows of virtual extensometer located +/- 0.6 mm above and below the crack (figure 4) is compared to the experimental profile issued from DIC, and the apparent friction coefficient is adjusted so that the computed and measured profiles best match.

Figure 5 illustrates the computations performed for various friction coefficients to find the one which best matches the experimental sliding displacement profile. When the most suitable friction coefficient (denoted below as "the apparent friction coefficient") is found, an elastic FE computation is made, using with this value, to estimate the effective stress intensity factor, using the G-theta method [44]. $\Delta K_{II,eff}$ can then be compared with the nominal stress intensity factor, issued from an elastic frictionless computation. For each of the tests, this inverse method was used for several crack lengths to obtain $\Delta K_{II,eff}$ and growth rate, to identify a Paris law in mode II.

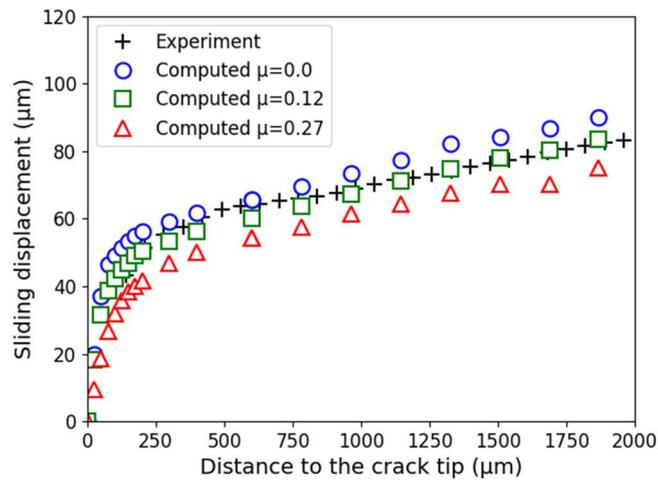

Figure 5: Comparison between the experimental and computed sliding displacement jump for various friction coefficient μ, taking crack tip plasticity into account, for $\Delta K_{II,nom}$ = 145.7 MPa√m

### III. Experimental and numerical results

#### III.1. Crack paths

Figure 6 shows the crack paths observed during each of the four tests, whose various "sequences" were described in table 2. The colored dotted lines superimposed on Fig. 6 indicate the end of some of the loading sequences.

Test #1 (figure 6(a)) had eleven loading sequences, to investigate the impact of the shear stress range and that of the superimposed biaxial compression on the crack path, and find out which combination has to be imposed to propagate the crack in pure mode II. The first four sequences did not include any biaxial compression, but only cyclic shear, with loading ranges, Δτ, along the crack decreasing stepwise from 440 to 200 MPa. Four branches, inclined by approximately ±45° initiated from the edges of the rectangular notch root, and grew in mode I, as predicted by usual bifurcation criteria. Then, for sequences 5 to 8, an equibiaxial compressive stress between 275 and 325 MPa was superimposed, and all cracks stopped growing, despite an increase in the cyclic shear stress range up to 550 MPa. For sequence 9, the shear loading range was sharply increased to 1000 MPa, and the biaxial compression reduced to 200 MPa, which caused a transition to shear mode growth, either by bifurcation (crack #2, #3 and #4), or by the macroscopic growth of a secondary branch (crack #1) in mode II, at the expense of the crack in mode I. These cracks propagated in mode II, without bifurcation, over distances ranging from 1.4 mm (crack #2) to 18.2 mm (crack #1). The shear stress range was then progressively reduced, without changing the biaxial compression. When $\Delta\tau$ reached 600 MPa (sequence 11), crack #1 branched, while the other 3 cracks propagated at a rate smaller than $10^{-9}$ m/cycle (near the resolution limit of the resolution the optical camera). When $\Delta\tau$ was raised again to 800 MPa (sequence 12), the two branches of crack #1 bifurcated back to mode II.

During test #2, a shear loading range $\Delta\tau$ = 1000 MPa with a biaxial compression of 200 MPa was applied. As observed in figure 6(b), two coplanar cracks initiated from the notch root, and propagated straight ahead over nearly 20 mm, without ever bifurcating into mode I. The shearing amplitude was thus sufficient to overcome the high contact pressure between crack faces, and promote shear-mode crack growth, in contradiction with most bifurcation criteria, which would have predicted an immediate bifurcation and propagation in pure mode I. The average growth rate of crack #1 is much larger during test 2 than that of crack #1 during sequence S9 of test 1, for the same loading conditions, probably because 1) the crack is longer, and thus the nominal $\Delta K_{II}$ is larger for test 2, and 2) the crack geometries are quite different, and the provided mechanical energy drives 4 cracks in test 1, but only two in test 2. The correlation of the measured crack growth rates with the effective $\Delta K_{II}$ presented below (Fig. 10) will rationalize these differences.

Test #3 (figure 6(c)) meant to document the lower part of the mode II Paris curve, the shear stress range was first 800MPa, and the biaxial compression 200 MPa, which led to the initiation of 4 branches at the edges of the circular notch, but these branches soon adopted a direction coplanar to the notch. They continued propagating in mode II when $\Delta\tau$ was raised to 900 MPa, as well as during sequences 3 and 4, during which $\Delta\tau$ was continuously reduced from 900 to 90 MPa, and from 900 to 540MPa, respectively, within 5000 cycles. For an equiaxial compression of 200 MPa, the cracks had no measurable growth when $\Delta\tau$ was below 600 MPa, in accordance with test #1.

Finally, test #4 on 58NCD13 steel (figure 6(d)) had seven sequences after mode I precracking. During sequences 1 and 2, shear stress ranges of 1400 MPa and 1600 MPa, respectively, with a biaxial compression of 300 MPa and 400 MPa, led to propagation in pure mode II of crack #1, while crack #2, at the other notch root, did not grow. Instead, 3 new branches initiated from the notch roots at ±45°, because the precracks length was insufficient to shield the notch roots and prevent the initiation of these branches. Thus, during sequence 3, equibiaxial traction was resumed, as an attempt to lengthen the coplanar cracks #1 and #2 and shield the other 3 branches, but this proved unsuccessful. For sequence 4, a shear loading range Δτ = 1600 MPa with an equibiaxial compression of 400 MPa was applied. A 200 MPa additional compressive stress along the x axis (figure 2), inclined by -45° relative to the notch, was superimposed, to prevent the opening of crack #3 and crack #5, which resulted in their arrest. Meanwhile, crack #4 grew in mode I over 0.95 mm, while crack #1 coplanar to the notch continued to grow in mode II, without bifurcation, over 2.3 mm, and crack #2 stopped growing, because of the shielding effect from crack #3 and crack #4. Then, for sequence 5, the shear stress range and biaxial compression were the same as during the previous sequence, except that the 200 MPa was now applied along the y axis, inclined by +45° relative to the notch plane, to prevent the opening of crack #4, which was successful. Meanwhile, the growth of crack #1 was not measurable. The last two sequences (6 & 7) were started with Δτ = 1600 MPa and an equibiaxial compression of 400 MPa, but Δτ was then continuously reduced to 1130 MPa within 650 cycles, and to 1390 MPa within 875 cycles, respectively. During both sequences, crack #1 continued propagating in mode II, over 645 and 875 μm, respectively.

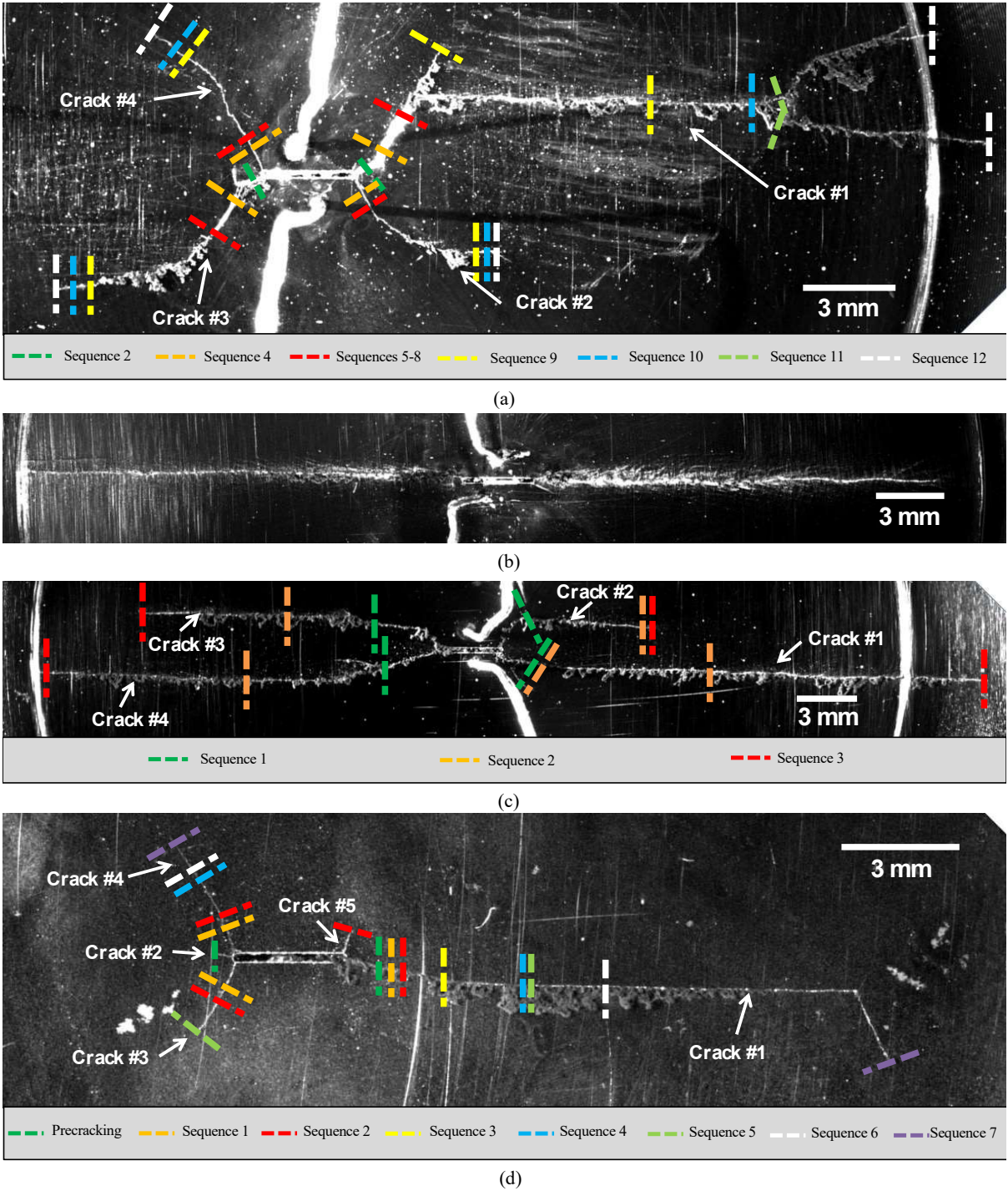

Figure 6: Crack paths issued from direct optical observation with all loading sequences: (a) test #1, (b) test #2, (c) test #3, and (d) test #4. The color bars indicate the end of the loading sequences.

Mode II propagation was associated with the formation of many branches starting from the crack flanks, at some distance behind the crack tip, and inclined by ±90° to ±110° with respect to the main crack growth direction (figure 7). For 16NCD13 steel, the length of these branches varied between 20 and 500 µm, while their length was less than 50 µm for 56NCD13 steel, independently of their position with respect to the main crack front. They stopped growing once the singular stress field around the main crack front moved away.

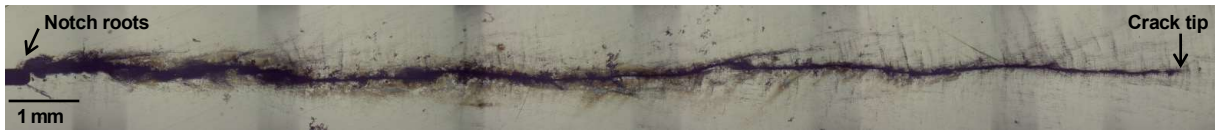

Figure 7: Optical observation of nearly transverse secondary branches in mode II for 16NCD13 steel (test #2)

**III.2. Observation of the fracture surfaces**

At the end of the tests, the specimens were broken open without damaging the fracture surfaces, by quenching it in liquid nitrogen, and striking it with a hammer to induce static brittle fracture. The images shown below correspond to crack #1 of tests #2 and #4, grown mostly in mode II. The fracture surfaces exhibit millimeter-long wear marks, as well as non-uniformly oxidized and mated areas, indicating an important friction between the crack lips, as observed in figure 8(a) (16NCD13 steel specimen test #2), captured using the back scattering electron signal in the SEM, to enhance the chemical contrast, or using color optical imaging (figure 8(e), for 58NCD13 steel specimen test #4).

The fracture surface of specimen #2 submitted to a constant shear stress range and a constant biaxial compression of 200 MPa over the whole test looks specially worn and oxidized near the notch root, where the crack faces were subjected to a larger number of shear cycles with a higher ΔCSD range, than near the final crack tip, where both the number of cycles and ΔCSD were smaller, as also reported by Bonniot et al. [20]. In addition, oxygen molecules might more easily access the crack faces, kept closed by normal compression, through the open notch. Such an oxidation gradient along the crack growth direction is not visible on specimen #4 (Fig. 8(e)), because of mode I precracking of the latter, over 3 mm, before applying 300 to 400 MPa biaxial compression. The relatively long and closed precrack did not constitute an easy path for oxygen. By contrast, the striking change in color between the side surfaces and the mid-thickness suggest an easier access of oxygen, and a higher degree of oxidation near the lateral sides of the contact area, as also reported for fretting tests with normal compression run by Baydoun et al. [19] and checked by Energy Dispersive Spectroscopy chemical analysis in the SEM, in the present study. Such a transverse gradient is also present in specimen #2, although less readily visible. To document the oxidation gradient along the thickness, the grey level of the BSE-SEM image (dark = 0, bright = 256) were averaged over 5x5 pixel-large moving windows, and plotted versus the distance from one side of the specimen to the other, for three distances from the notch root: 1.7, 7.1, and 15.4 mm (figures 8(b), (c), (d), respectively). Bright areas represent the bare steel, while the darker areas are oxidized. The first two profiles clearly reveal darker outer surfaces, through which oxygen molecules can access and react with the steel surfaces, freshly stripped by cyclic shear, while the mid-thickness area, where oxygen did not access, due to important contact pressure, looks brighter. Finally, at 15.5 mm from the notch root, the grey level remains constant, but high (≈200) over the whole thickness, which implies very limited oxidation at that fast growth stage. These longitudinal and transverse oxidation gradients have direct consequences in terms of crack growth kinetics, since oxidized metal debris are considered by tribologists as a "third body" that prevents direct contact between bare metallic surfaces of both crack lips, and shields the adhesive forces, thus reducing the friction coefficient and raising the effective $\Delta K_{II}$.

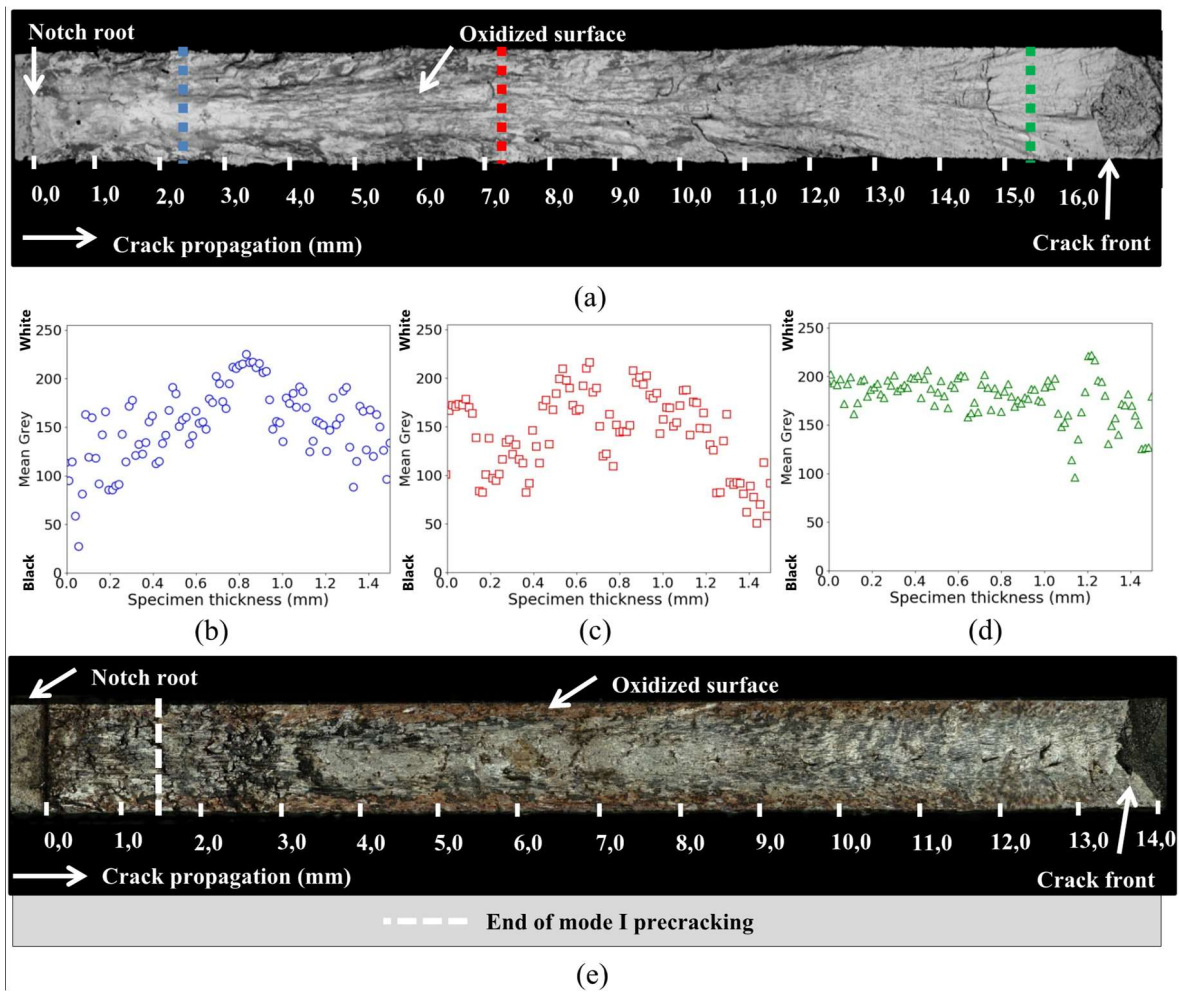

Figure 8: Fracture surfaces after shear mode crack propagation: (a) 16NCD13 steel (test #2) observed using back scattering electron signal in the SEM to enhance chemical contrast, (b-d) Mean grey level of BSE-SEM image (b) at 1.7 mm from the notch root, (c) at 7.1 mm from the notch root, (d)) at 15.4 mm from the notch root, and (e) Fracture surface of 58NCD13 steel (test #4) using colour optical observation to highlight oxidation gradients.

### III.3. Mode II Crack growth kinetics

The application of the inverse numerical method described above on crack #1 for each of the three tests run on both steels allowed the computation of $\Delta K_{II,eff}$ at various stages. These cracks were chosen because they propagated the most in mode II. The evolution of $\Delta K_{II,eff}$ with crack length during test #2, run under constant shear stress range plus biaxial compression is compared to that of $\Delta K_{II,nom}$ on Figure 9a, on which the difference,

$$\Delta K_{II,f} = \Delta K_{II,nom} - \Delta K_{II,eff}, \qquad (6)$$

is also plotted.

When the crack is only 2.5mm long, the nominal and effective SIFs are nearly similar, which suppose that extrinsic mechanisms dissipate little energy at that stage. However, the difference between the nominal and effective SIFs (that is $\Delta K_{II,f}$) rises during crack propagation, which means that the energy dissipated by friction increases with the crack length.

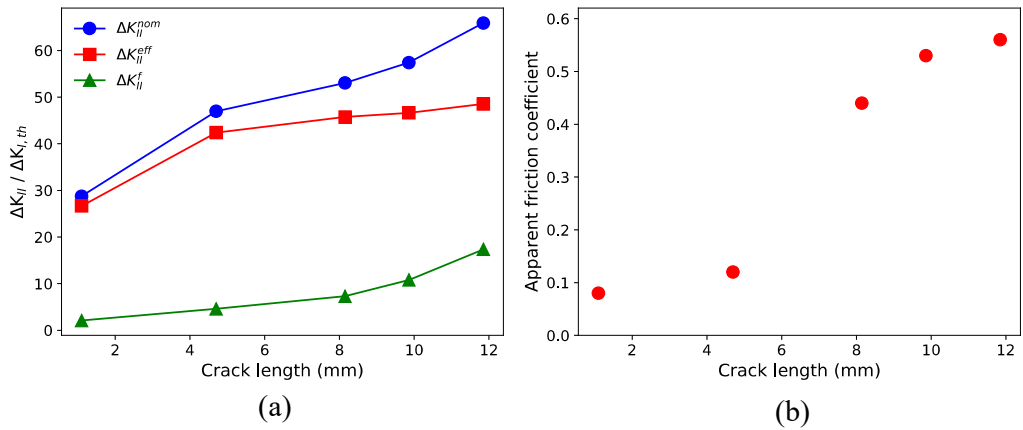

**Figure 9: a) Evolution of the stress intensity factors under a constant cyclic shear range + biaxial-compression (test #2) and b) corresponding evolution of the apparent friction coefficient**

The inverse analysis of the measured crack sliding displacement profiles also provides an apparent friction coefficient, µ, for each crack length. The evolution of µ as a function of crack length during test #2 is plotted on figure 9b. A strong and quasi-linear rise of the apparent friction coefficient during crack propagation -from approximately 0.08 to 0.56- is observed, and will be discussed below.

The mode II growth rates measured for crack #1 during each of the three tests are plotted as a function of $\frac{\Delta K_{II,eff}}{\Delta K_{I,th}}$ (rather than versus $\Delta K_{II,eff}$ for confidentiality reasons) on figure 10a. The data-points, which cover four decades of crack growth rate ($5.10^{-9}$ to $5.10^{-5}$ m/cycle), align well on a single, intrinsic curve, which can be described by a Paris equation as:

$$\frac{da}{dN} = C(\Delta K_{II}^{eff})^m, \qquad (7)$$

where C and m are material coefficients (not provided, for industrial confidentiality reasons).

Mode I crack growth kinetics were obtained for the two steels at three different R ratios (-1, 0.05 and 0.5), using CT specimens. The influence of the R ratio was very limited, at least in the Paris regime, within which closure effects probably vanished at R = 0.5, so that $\Delta K_{I,eff} \approx \Delta K_{I,nom}$. The corresponding mode I crack growth kinetics are thus compared to the da/dN- $\Delta K_{II,eff}/\Delta K_{I,th}$ curves, on figure 10b. For the same effective $\Delta K$, the crack growth rate is higher in 58NCD13 than in 16NCD13, whatever the propagation mode. The Paris exponents of mode I and mode II kinetics are nearly the same.

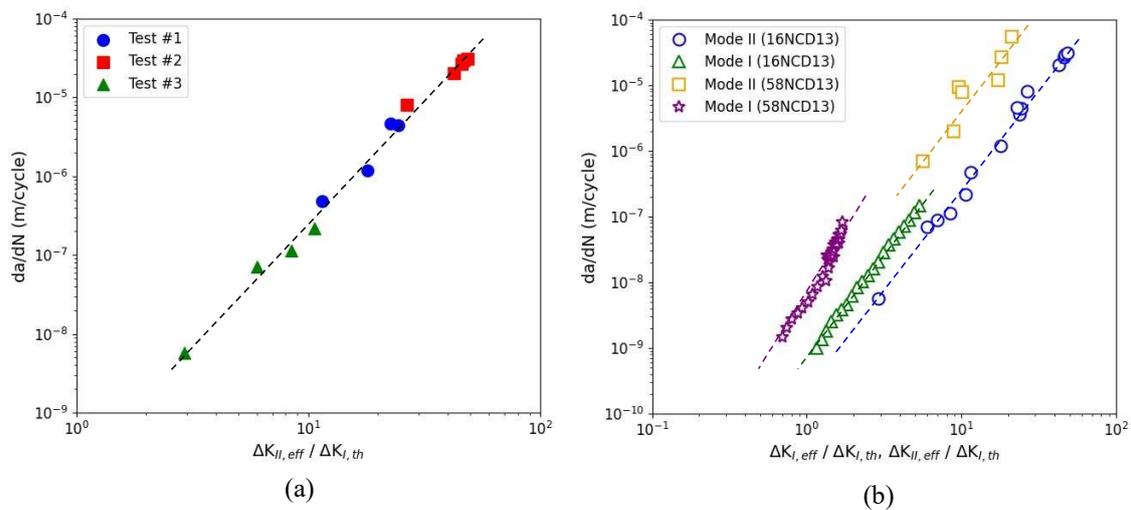

Figure 10: a) Mode II crack growth rate as a function of $\frac{\Delta K_{II,eff}}{\Delta K_{Ith}}$ for each of the three tests on 16NCD13 steel b) Comparison between mode I and mode II kinetics for 16NCD13 and 58NCD13 steels

### IV.    Discussion

**IV.1 Mode II vs Mode I crack growth kinetics**

In most studies comparing closure-corrected mode I and friction-corrected mode II [22,25,45] crack growth kinetics, or mode I and friction-corrected mode III kinetics [35,36] in various steels or titanium alloys, a higher exponent for the Paris law was found in mode II (or mode III) than in mode I, so that opening and shear mode curves intersect, and for a high $\Delta K_{eff}$, above the intersection point, shear mode propagation is faster than mode I, and inversely for low $\Delta K_{eff}$. In such cases, the maximum crack growth rate criterion [45] predicts crack paths that are consistent with experimental observations. For 16NCD13 and 58NCD13 steels, the results presented in figure 10b suggest nearly similar Paris exponents for mode I and mode II crack growth kinetics. However, the growth rate range within which both mode I and mode II data are available is still limited, and more tests would be necessary to document the lower part of the Paris regime in mode II. Note however that Bonniot [46] also found quasi-superimposed mode I and mode II crack growth kinetics for a rail steel. In the present study, the extrinsic effect of normal compression on the crack driving force was taken into account through the use of $\Delta K_{II,eff}$. It cannot be excluded though, that the biaxial compression has a moderating influence on the damage mechanisms ahead of the tip, and thus on the crack growth rate.

**IV.2 Crack face friction and its evolution**

Crack face friction plays an essential role in mode II fatigue crack growth, even when no normal compression is applied, because of crack face roughness, which leads to asperities contact and interlocking [23,24]. Coulomb's friction law, which cannot capture this effect unless the actual roughness is considered, predicts that crack face friction should be enhanced when normal compression is applied. However, compression may also favor the trapping of wear debris, restrict the access of oxygen between contacting surfaces, thus changing -and sometimes actually reducing- the friction coefficient, and favor adhesive rather than abrasive wear [19]. Besides, as discussed by Bonniot et al. [20] based on Archard's wear law [47], for a fixed CSD, normal compression should enhance wear, but on the other hand, it should also reduce $\Delta K_{II,eff}$ and thus the CSD, so that the resulting effect is not a systematic enhancement of crack face wear, but depends on the relative magnitude of the applied shear stress range, and the product of the friction coefficient by the compressive stress. During test #2, run under constant shear stress range and static biaxial compression, a quasi-linear and significant rise of the apparent friction coefficient was observed during crack propagation (see Fig. 9b). The observation of the different shades of grey on the BSE-SEM images of the fracture surfaces shown in figure 8 suggests an explanation based on oxidation gradients. The crack surfaces look more oxidized near the notch root than within the last two to three millimeters of crack growth. This might partly be due to the rise in crack growth rate associated with the rising $\Delta K_{II,eff}$, (see Fig. 9a) and thus the decrease in the number of contact and friction cycles undergone by points of the crack faces with their distance to the notch root. Besides, the notch, which does not close completely under the applied compression could be a source of oxygen flow from the environment, into the crack. It is nonetheless fair to note that, due to its roughness, the crack cannot be considered as fully closed, because sliding of asperities can locally induce some opening displacements. However, the access of oxygen is certainly more difficult under normal compression. Differences in the degree of oxidation also exist between the lateral surfaces, from which external oxygen might flow, and the mid-thickness region. An oxidation gradient at the external border of contacting surfaces during fretting tests was reported by Baydoun et al [19], modelled in terms of oxygen diffusion and reaction with the metal, and associated with a transition from abrasive wear, and low friction coefficient, in the external, oxidized areas, to adhesive wear, and high friction coefficient, in the metal-on-metal central contact area. Anyway, the oxide layer between crack faces, also called "third body" by Godet [48] is known to prevent direct contact between fresh metallic surfaces, thus decreasing adhesive forces, and to accommodate the sliding displacement jump, thus reducing the friction coefficient. Thus, the presence of this third body, more important in the oxidized zones, can also partly explain the observed evolution of the apparent friction coefficient during crack propagation. Finally, the term "apparent" should be stressed, because this parameter is issued from the best fit of the measured CSD profiles, assuming a uniform friction coefficient. The fact that this "average" value increased from 0.08 to 0.56 during crack growth suggests even more pronounced local gradients.

**IV.3 Crack growth competition**

Murakami & Takahashi [31] illustrated the crack growth competition between a crack propagating in mode II, and a branch initiated along the main crack flank, and propagating in mode I. The presence of this branch reduces $\Delta K_{II,eff}$ at the main crack tip, which propagates at the expense of the branch, despite the shielding effect from the latter. As explained by Doquet & Frelat [37], and Dubourg & al. [30], microcracks have a shielding effect on the main crack, reducing its effective SIFs and growth rate. Moreover, crack bifurcation will not occur in the same conditions for a single crack and a crack in a network [30]. Consequently, the presence of secondary cracks impacts directly the crack path and crack growth rate. Doquet and Frelat [37] developed a 2D FE model to analyze -with or without crack face friction- the shielding effect of a secondary branch of length *s*, orthogonal to the flank of a smooth crack loaded in mode II, at a distance *b* behind its tip. $\Delta K_I$ for this branch, as well as $\Delta K_{II}$ for the main crack, were computed as a function of b/s, and the simultaneous and competitive growth of the main crack, in mode II and the branch crack, in mode I (neglecting the smaller mode II component that it also undergoes) were simulated step by step, using experimental kinetic data. Branching from the crack flank rather than continued coplanar growth or even kinking from the main crack tip were shown to occur at low $\Delta K_{II,eff}$. In the present study, the same type of computation is made, but using a 3D FE model of cruciform specimens, and taking into account 1) the inclination angle of the branches, which was found to vary between 90 and 110°, 2) their multiplicity and 3) the influence of biaxial compression.

*IV. 3.1. Branch crack initiation*

As suggested by Kfouri [49] and Doquet & Frelat [37], such branches can initiate from the crack tip, along secondary slip bands, since, in elasticity, their orientation is not far from the second angular extremum of the shear stress:

$$\sigma_{r\theta} = \frac{K_{II}}{2\sqrt{2\pi r}} \cos\left(\frac{\theta}{2}\right)(3\cos\theta - 1), \tag{8}$$

at ±124°, while the highest peak value, responsible for the primary slip band ahead of the crack tip is at θ = 0°. Such an argument assumes that the branches initiate from the tip and then are left behind by the propagating main crack. However, if they do initiate from the flank (θ = ±180°), due to an opening stress, their orientation, nearly orthogonal to the flank of the main crack is consistent with the peak values of $\sigma_{rr}$: θ = -180° (lower crack flank) when $K_{II}$ > 0 and: θ = 180° (upper crack flank) when $K_{II}$ < 0. Anyway, once initiated, these branches undergo cyclic mode I [37,49], whose amplitude rises with $\Delta K_{IIeff}$ at the main crack tip, and decreases with their distance to this tip. This can allow them to grow beyond the first grain boundary. Indeed, in the present study, some of these branches reached 500 µm in length, much longer than the mean size of lamellae colonies sharing the same crystallographic orientation. Their growth (under the influence of biaxial compression, which was not considered in [37,49]), and their shielding effect on the main crack, will be further discussed below.

*IV.3.2 Effect of the inclination angle of the branches –*

A 3D FE model was developed, in which a secondary branch, inclined by 90° or 110° to the main crack flank was inserted at a distance *b* behind the main crack tip (figure 11). The length of the main crack was 4.7 mm and the length of the branch s = 0.10 mm. A cyclic shear stress Δτ = 1000 MPa and a static equibiaxial compression of 200 MPa where applied, like in test #2. Several stationary elastic computations were performed for different values of *b,* in order to study the evolution of the shielding effect and the loading condition of the branch, as a function of the b/s ratio. Coulomb's friction law was applied between both the main crack and branch crack lips, with a friction coefficient µ = 0.12 (equal to the apparent friction coefficient obtained by inverse method for this crack length during test #2). A complete cycle was computed, so as to derive $\Delta K_I$ and $\Delta K_{II}$ at mid-thickness and at the surface to study if there is an evolution of the loading and of the shielding effect along the crack front, and the results were found to be similar.

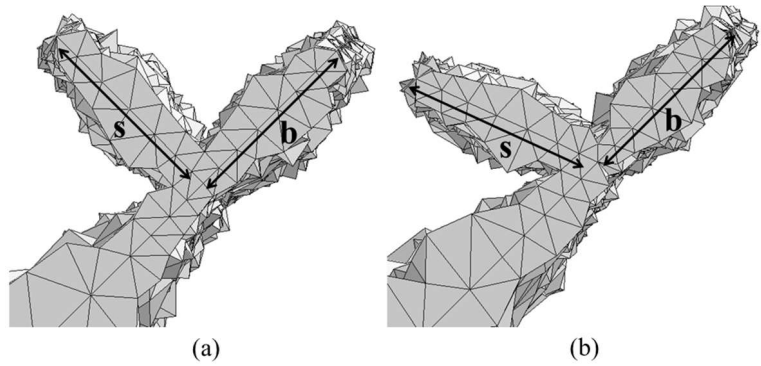

**Figure 11: Meshes used for 3D FE computations of the SIFs with a secondary branch inclined by: (a) 90° and (b) 110° relative to the main crack submitted to cyclic shear plus biaxial compression.**

Figure 12 compares the evolutions with b/s of $\Delta K_I$ and $\Delta K_{II}$ at the tip of a branch inclined by 90° or 110°. The 90° oriented branch (figure 12(a)) is mainly driven by mode I. It is also loaded in mode II, especially when it is close to the main crack front, although mode I is still largely predominant. By contrast, the 110° inclined branch is nearly equally loaded in mode I and mode II (figure 12(b)). This branch is subjected to a higher $\Delta K_{II}$, but a smaller $\Delta K_I$ than the one oriented at 90°. In order to understand which of the two orientations is the most critical for the crack growth competition with the coplanar crack, figure 12(c) compares the evolutions with b/s of the energy release rate amplitude $\Delta G$ on the branches. $\Delta G$ is only slightly higher for the orthogonal branch. However, as mentioned above, the presence of $T_y < 0$ should favor this orientation.

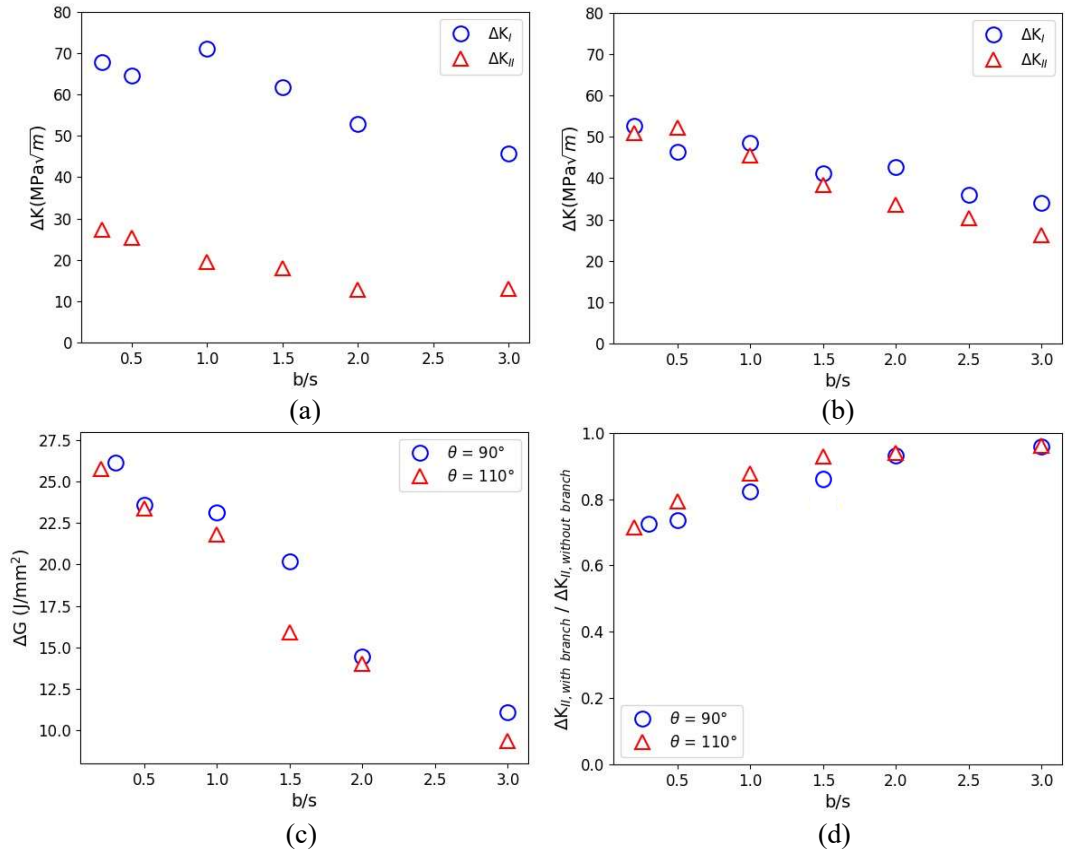

**Figure 1~~1~~2**: Evolutions with b/s of $\Delta K_I$ and $\Delta K_{II}$ at the tip of a branch inclined by: (a) 90°, (b) 110°, behind the tip of a crack loaded in mode II. Evolution with b/s of $\Delta G$ at the tip of the branch, and (d) Influence of the branch inclination on the shielding effect of the main crack

In order to study the shielding effect of such branches, the ratio between $\Delta K_{II,eff}$ of the main crack with the branch to its value without the branch is plotted versus b/s on figure 12(d). A slight influence of the inclination angle of the branch on its shielding effect can be noted. A crack oriented at 110° has stronger shielding effect than an orthogonal branch for a low b/s ratio (20% to 30% reduction of $\Delta K_{II,eff}$ for b/s< 0.5), however its shielding effect vanishes much faster, as b/s rises. This implies that the orthogonal branch is definitely more critical for the growth competition with the main crack. Only this orientation will thus be considered in the following analyses, and the mode II component on such branch will be neglected.

*IV.3.3. Influence of multiple orthogonal branches –*

To analyze the influence of the multiplicity of secondary branches on the evolutions of the SIFs at the main crack front, 3 orthogonal branches were inserted in the previous 3D FE model. The length of each branch was s = 0.10 mm, and the distance between them was also 0.10 mm. Several stationary elastic computations were performed for different values of *b* – the distance between the main crack tip and the nearest branch- in order to study the evolution of the shielding effect and the loading condition of the branch, as a function of b/s. Figure 13 shows the evolution of $\Delta K_I$ for the branch nearest to the main crack tip with, or without, the other two branches. The latter two reduce $\Delta K_I$ for the nearest branch. Thus, this branch is less likely to propagate, because of shielding effects from other branches. Consequently, the main crack has a better opportunity to win the crack growth competition. The influence of the branch cracks multiplicity on the shielding effect for the main crack is illustrated on figure 13(b). When b/s is lower than 1, that is: when the nearest branch is relatively close to the main tip, no significant effect of the multiplicity of secondary branches is found. However, multiple branch cracks provide a more persistent shielding effect on the main crack tip when b/s increases, than a single one.

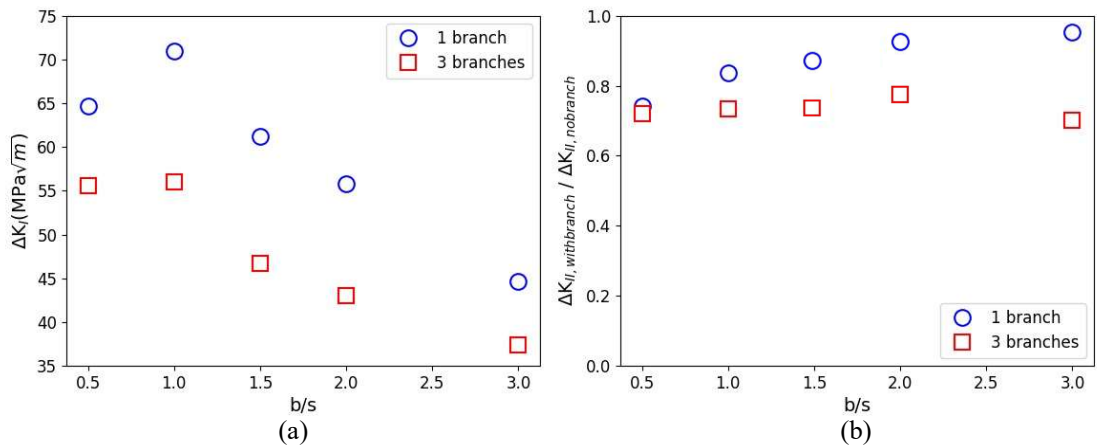

**Figure 13:** (a) Evolution with b/s of $\Delta K_I$ for the branch nearest to the main crack tip with or without surrounding branches, and (b) Shielding effect of the branches on the main crack: 1 branch vs 3 branches

*IV.3.4. Influence of a biaxial compression on the kinetic competition –*

To clarify the role of the biaxial compression on the kinetic competition, different equibiaxial compressions (0 MPa, 200 MPa and 400 MPa) were applied on the 3D FE model with a 4.7 mm long main crack, a single orthogonal branch crack with an initial b/s = 1, and $\Delta\tau$ = 1000 MPa, as in test #2. An iterative elastic computation was run. At each step, $\Delta K_I$ at the tip of the branch, and $\Delta K_{IIeff}$ at the main crack tip were computed, and their respective growth rates were deduced, based on the mode I and mode II kinetics presented in figure 10b. Figure 14 plots the predicted evolutions of b/s with the number of propagation cycle. Note that an increase in b/s means that the main crack propagates faster in mode II than the branch, which is left more and more behind the main tip, and thus gets unloaded, as illustrated in figure 15(a). In all simulated cases, b/s rises, which means that for this $\Delta\tau$, mode II crack growth is predicted, rather than branching from the crack flank. However, the higher the biaxial compression, the faster b/s rises, as observed on figure 15, in spite of the friction-induced reduction of $\Delta K_{IIeff}$ visible on figure 15(b). The increase of $\Delta K_{IIeff}$ with b/s, when the equibiaxial compression is 0 MPa, is due to the progressive decrease of the shielding effect resulting from the presence of the branch, to a large extent. However, when a biaxial compression is applied, $\Delta K_{IIeff}$ hardly evolves with b/s, especially at 400 MPa, which indicates a reduction of the influence of the branch on the main crack when biaxial compression increases.

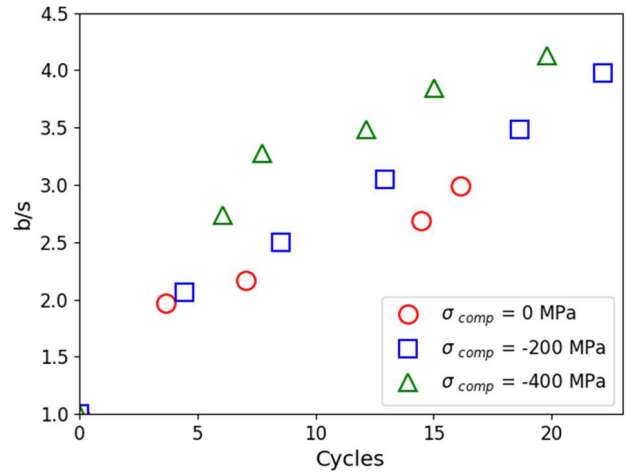

Figure 2-14: Competition between the main and branch crack for various biaxial compressions

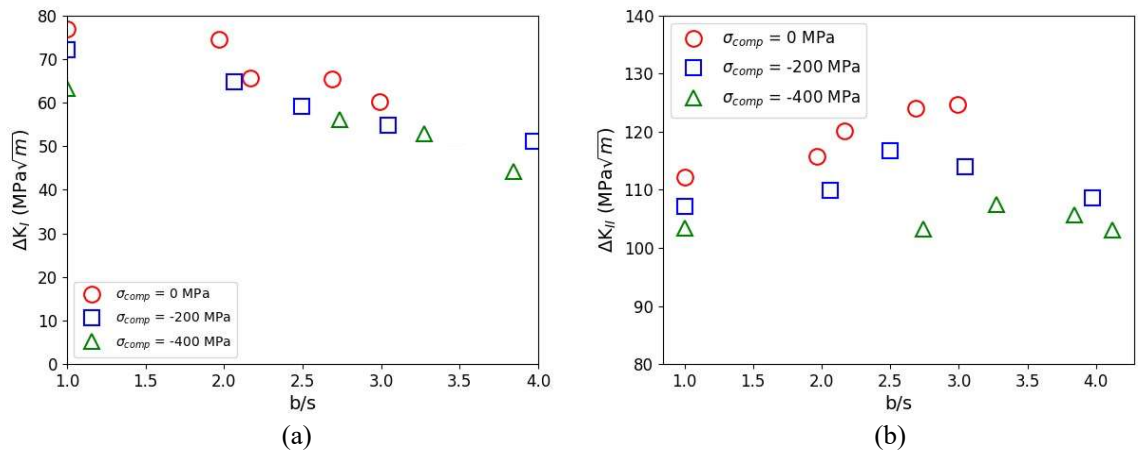

Figure 15: (a) evolution of $\Delta K_I$ on the branch and (b) evolution of $\Delta K_{II}$ on the main crack for various biaxial compressions, as a function of b/s

To analyze separately the respective influences of the two non-singular stresses $T_x$ and $T_y$ on the growth competition between an orthogonal branch initiated from the flank and the main crack loaded in reversed mode II, 3D finite element calculations were run on a large plate with a central crack and a normal branch with a b/s ratio of 1 (figure 16(a)). The computations were performed with a friction coefficient of 0.1, between the crack faces, and $\Delta K_{II} = 36\ \text{MPa}\sqrt{\text{m}}$. On the one hand, a compression $T_y$ perpendicular to the main crack enhances friction and thus reduces its sliding displacement and $\Delta K_{IIeff}$ at the main crack tip. It also reduces the opening displacement of the orthogonal branch, and thus $\Delta K_I$ at its tip (figure 16(b)). However, the reduction of $\Delta K_I$ for the branch (-32% for $T_y$ = - 400 MPa) is much stronger than the reduction of $\Delta K_{IIeff}$ at the main crack tip (-13% for $T_y$ = - 400 MPa). Paradoxically, the normal compression thus favors continued coplanar growth in mode II, rather than the development of the orthogonal branch. On the other hand, the compressive stress, $T_x$, applied alone reduces $\Delta K_I$ on the secondary branch to a much larger extent, and slightly increases $\Delta K_{IIeff}$ at the main crack tip. Such an increase is not observed at the other crack tip, where no branch is present. The increase in $\Delta K_{IIeff}$ is thus an effect of the presence of the branch. By reducing the opening displacement of the secondary branch, Tx reduces its shielding effect on the main crack, this explaining the increase of $\Delta K_{IIeff}$. A negative $T_x$ thus clearly, promotes coplanar fatigue crack growth in mode II, as experimentally observed by Otsuka and al. [22]. When equibiaxial compression is applied (Tx = Ty), the reduction of $\Delta K_I$ on the secondary branch (-41% for $T_x = T_y$ = - 400 MPa) dominates over the reduction of $\Delta K_{IIeff}$ at the main crack tip (-9% for $T_x = T_y$ = - 400 MPa) (figure 16(b)), so that mode II crack growth and an arrest of the branch should be promoted.

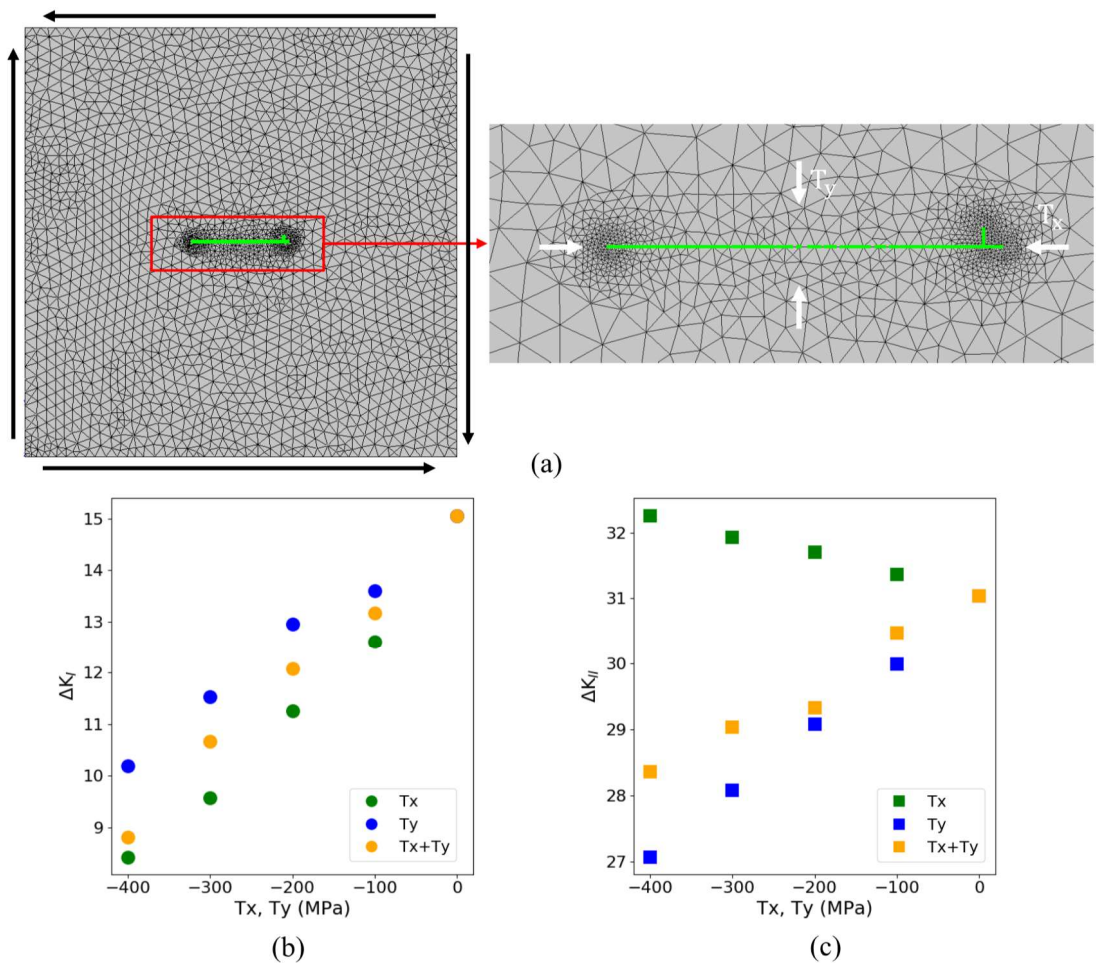

**Figure 16: (a) Mesh used for the computation of the SIFs for a crack loaded in mode II with an orthogonal branch with b/s=1, in a large plate, (b) evolution of $\Delta K_I$ on the branch and (c) evolution of $\Delta K_{II,eff}$ on the main crack for various T-stress components applied alone or simultaneously.**

Therefore, the reasons for the coplanar growth of the main crack are the biaxial compression -which reduces $\Delta K_I$ of the branches-, and the important shear stress range which allows the main crack to grow in mode II, even though the normal compression reduces $\Delta K_{II,eff}$. Gates & Fatemi [50] suggest that the effective driving force in mode II increases faster with the loading range than the attenuation due to crack face interaction. A high loading range probably enhances the wear of crack faces asperities and thus the effective shear-mode crack driving force.

## V. Conclusions

Mode II fatigue crack growth under reversed shear loading and superimposed static biaxial compression was investigated in 16NCD13 bearing steel, as well as in 58NCD13 steel, representative of the surface layer of carburized bearings, in terms of carbon content, using cruciform specimens with a center crack.

Many aborted branches, quasi-orthogonal to the main crack, were observed along the crack face. The influence of their inclination angle and multiplicity on the stress intensity factors at their tip, as well as their shielding effect on the main crack were analyzed by FE simulations. While the branches inclined by 110° were found to undergo nearly equal mode I and mode II loading ranges, those inclined by 90° undergo mostly cyclic mode I. When several such branches are present, their shielding effect on the main crack is stronger than that of a single branch. Biaxial compression was shown to hinder the growth of these branches and to favor coplanar mode II crack growth. The role of the compressive stresses applied parallel and normal to the main crack plane (Tx and Ty, respectively) was analyzed separately.

Normal compression hinders the access of oxygen into the closed crack, inducing a gradient in crack face oxidation, and a transition from abrasive wear, near the side surfaces, to adhesive wear, at mid thickness. It reduces $\Delta K_{II,eff}$, but also $\Delta K_I$ at the tip of orthogonal branches. This oxidation gradient also leads to an evolution of the apparent friction coefficient during the crack propagation. Parallel compression reduces $\Delta K_I$ at the tip of orthogonal branches, and thus their shielding effect on the main crack, finally increasing $\Delta K_{II,eff}$. It makes mode II coplanar growth more stable.

The crack face sliding displacement profiles measured by DIC were used to derive $\Delta K_{II,eff}$ at the main crack tip, using elastic-plastic FE simulations with crack face friction, by an inverse method. Friction-corrected crack growth kinetics were thus obtained for mode II crack growth in both steels, and compared to the crack growth kinetics in mode I. Contrary to previous studies on other materials, the Paris exponents of mode I and mode II kinetics were found here to be nearly the same.

**Acknowledgements**

The authors would like to acknowledge Airbus Helicopters for its financial support.

**Appendix – Identification of the constitutive equations**

Strain-controlled uniaxial traction-compression tests were made, and the strain-stress curved were recorded. To establish elastic-plastic constitutive equations, a model from Lemaitre & Chaboche [38] was used:

$$f(\underline{\sigma}, \underline{X}, R) = J(\underline{\sigma} - \underline{X}) - \sigma_y - R, \tag{A1}$$

where f is the yield function, $\underline{X}$ is the kinematic hardening tensor variable, R is the isotropic hardening scalar variable, and $\sigma_y$ represents the yield stress. The hardening variables are defined by:

$$\begin{cases} \underline{\dot{X}} = \frac{2}{3} C \underline{\dot{\varepsilon}}^p - D \underline{X} \dot{p} \\ \dot{R} = b(Q - R)\dot{p} \\ \underline{\dot{\varepsilon}}^p = \dot{\lambda} \frac{\partial f}{\partial \underline{\sigma}} \end{cases}, \tag{A2}$$

with p the accumulated plastic strain, $\lambda$ the plastic multiplier and C, D, b and Q are materials coefficients.

The material exhibits cyclic softening, captured by a negative value of parameter Q. Two kinematic hardening variables ($X_1$ and $X_2$) were necessary to describe accurately the shape of stress-strain loops. The determination of each parameter was performed via a minimization of the difference between experimental and computed strain-stress loops. The measured and simulated loops are compared on figure A.1. For industrial confidentiality reasons, the stress has been normalized

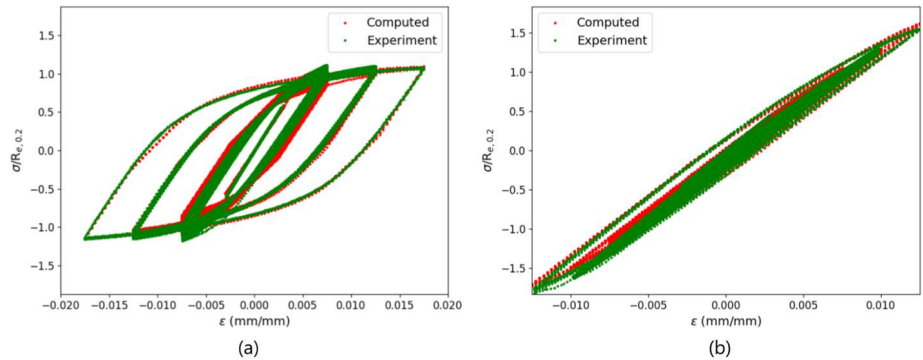

Figure A.1: Measured and simulated stress-strain curves for strain-controlled push-pull tests (a) on 16NCD13 steel, and (b) on 58NCD13 steel. For industrial confidentiality reasons, the stress has been normalized.